\documentclass{PoS}

\title{First Results on Hadron Spectroscopy at COMPASS}
\ShortTitle{First Results on Hadron Spectroscopy at COMPASS}

\author{\speaker{Frank Nerling}\thanks{on behalf of the COMPASS collaboration}\\
        Physikalisches Institut, Albert-Ludwigs-Universit\"at Freiburg\\
        E-mail: \email{nerling@cern.ch}}

\abstract{
The COMPASS fixed-target experiment at the CERN SPS is dedicated to the study of hadron
structure and dynamics. One goal of the physics programme using hadron
beams is the search for new states, in particular the search for $J^{PC}$
exotic states and glueballs. After a short pilot run in 2004 (190
GeV/c $\pi^{-}$ beam, lead target), we started our hadron spectroscopy
programme in 2008 by collecting unprecedented statistics using 190 GeV/c negative
hadron beams on a liquid hydrogen target. A similar amount of
data with 190 GeV/c positive hadron beams has been taken in 2009, as well
as some data (negative beam) on nuclear targets. 
As a first result the observation of a significant $J^{PC}$ spin-exotic signal 
in the 2004 data -- consistent with the disputed $\pi_1(1600)$ -- was recently 
published.
Our spectrometer features good coverage by electromagnetic calorimetry,
crucial for the detection of final states involving $\pi^0$, $\eta$ or $\eta'$,
and the 2008/09 data provide an excellent opportunity for the simultaneous
observation of new states in different decay modes. 
The diffractively produced $(3\pi)^{-}$ system for example can be
studied in $\pi^{-}\pi^{+}\pi^{-}$ and $\pi^{-}\pi^{0}\pi^{0}$ final
states, respectively. Observation of new states in both modes provides 
important consistency checks within the same experiment as the reconstruction of charged 
and neutral modes rely on completely different parts of the apparatus.
We present the first results and give an overview of the status on
various ongoing analyses of the 2008/09 data.
}
\FullConference{35th International Conference of High Energy Physics - ICHEP2010,\\
		July 22-28, 2010 \\
                Paris France}
\begin{document}
\vspace{-0.3cm}
\section{Introduction}
\vspace{-0.3cm}
%%\noindent
The existence of exotic states beyond the constituent quark model
(CQM) has been speculated about almost since the introduction of 
colour~\cite{Jaffe:1976,Barnes:1983}. 
Due to the self-coupling of gluons via the colour-charge, so-called 
hybrid mesons ($q\bar{q}$ states with an admixture of gluons) and 
glueballs (purely gluonic states without quark content) are allowed 
within Quantum Chromodynamics while they are beyond the CQM. 
The lowest mass glueball candidate is predicted
%%~\cite{cmcneile:2006} 
to have scalar quantum numbers of spin, parity and charge conjugation $J^{PC}= 0^{++}$, and a 
mass of $\sim 1.7\,{\rm GeV/c}^2$. 
Even though candidates have been reported by the Crystal Barrel and
the WA102 experiments, mixing with ordinary isoscalar mesons makes the 
interpretation difficult. 
\begin{figure}[tp!]
  \begin{minipage}[h]{.59\textwidth}
    \begin{center}
\vspace{-0.5cm}
     \includegraphics[clip, trim= 45 60 60 90,width=1.0\linewidth]{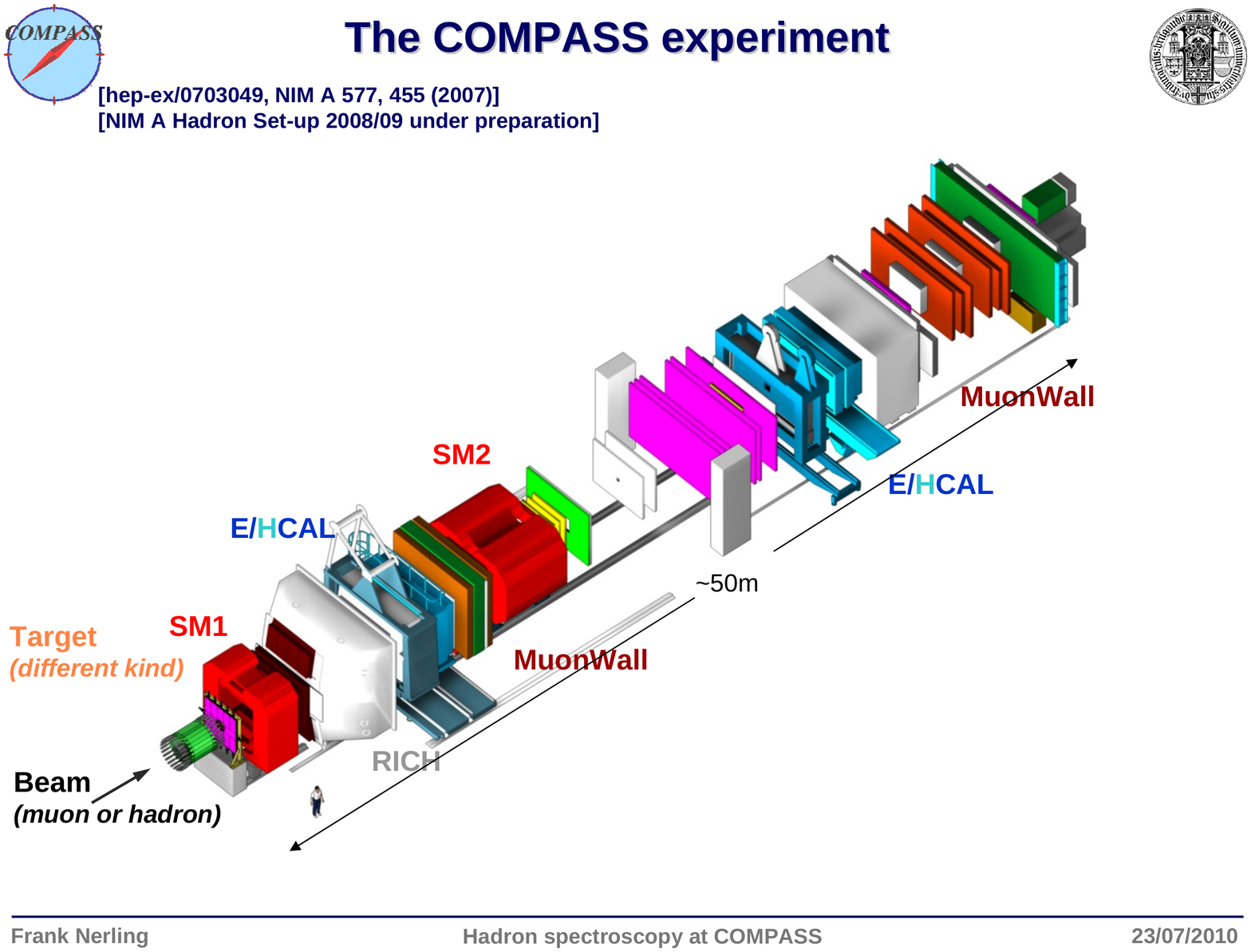}
    \end{center}
  \end{minipage}
  \hfill
  \begin{minipage}[h]{.39\textwidth}
    \begin{center}
\vspace{-0.5cm}
      \includegraphics[clip,trim= 20 0 0 0,width=0.8\linewidth,
       angle=0]{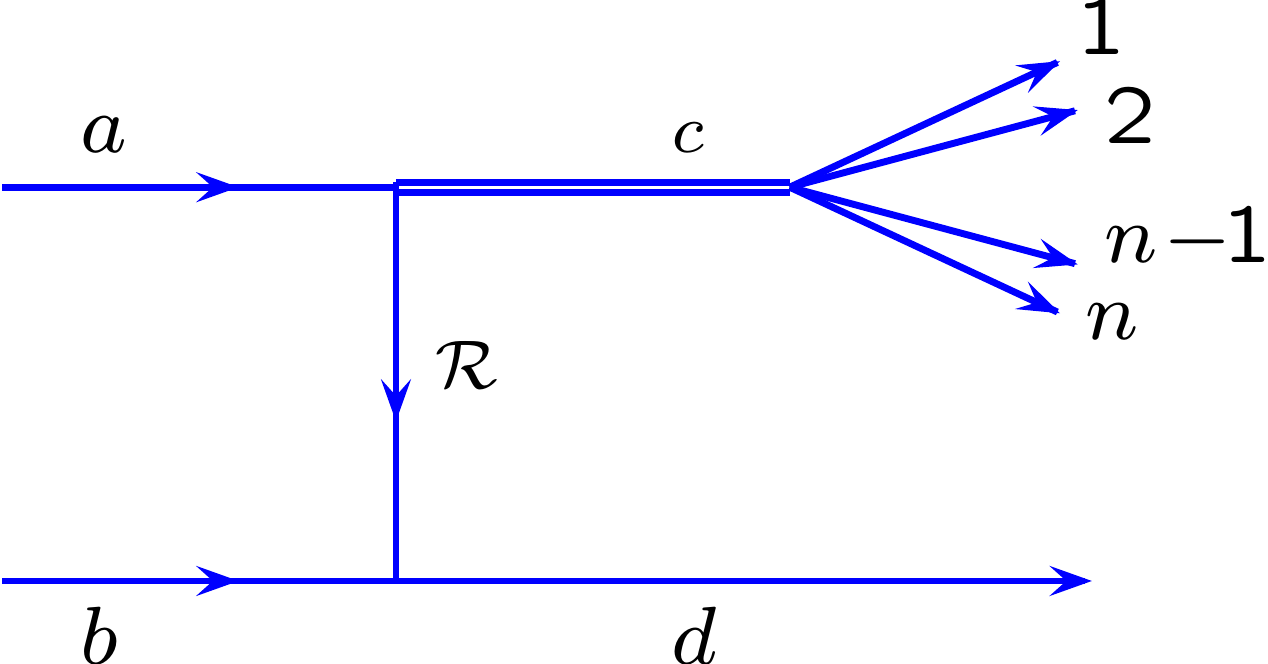}
\vspace{0.3cm}
      \includegraphics[clip,trim= 17 10 3 -10,width=0.8\linewidth,
     angle=0]{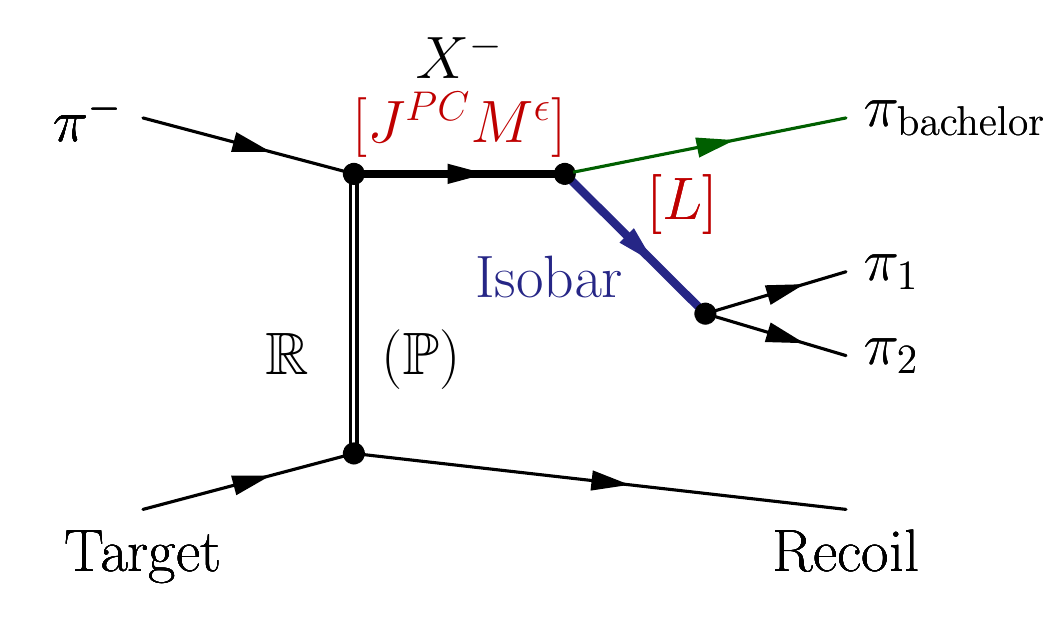}
    \end{center}
  \end{minipage}
      \caption{Left: Sketch of the two-stage COMPASS spectrometer.
               Right: \textit{(top)} Diffractive meson production via $t$-channel Reggeon exchange: The beam particle $a$ is excited to 
                     a resonance $c$ subsequently decaying into $n$ mesons, the target stays intact. 
                     \textit{(bottom)} Diffractive dissociation into 3$\pi$ final states as described in the isobar model: The produced
                     resonance $X^{-}$ with quantum numbers $J^{PC}M^\epsilon$ decays into an isobar with spin $S$ and relative orbital
                     angular momentum $L$ with respect to the $\pi_{\rm bachelor}$. The isobar subsequently decays into two pions.
                     At high energies, the Pomeron is the dominant Regge-trajectory.~~~~~~~~~~~~~~~~~~~~~~~~~~~~~~~~~~~~~~~~~~~~~~~~~~~~~~~~~~~~~~~~~~~~~~~~~~}
       \label{fig:diffrProd_Spectro} 
\vspace{-0.5cm}
\end{figure}
Several light hybrids, on the other hand, are predicted to have exotic 
$J^{PC}$ quantum numbers and are thus promising candidates in the
search for physics beyond the CQM. The hybrid candidate lowest in mass 
for example is predicted
%%~\cite{Morningstar:2004} 
to have a mass between 1.3 and 2.2\,GeV/c$^2$ and exotic quantum numbers $J^{PC}=1^{-+}$, 
not attainable by ordinary $q\bar{q}$ states.
Two experimentally observed $1^{-+}$ hybrid candidates in the light-quark sector have been reported in different decay 
channels so far, the $\pi_1(1400)$ mainly seen in $\eta\pi$ decays by e.g. E852~\cite{E852}, VES\cite{Beladidze:1993}, and Crystal 
Barrel~\cite{CB}, and the $\pi_1(1600)$ observed by both E852 and VES in the decay channels $\rho\pi$~\cite{Adams:1998,Khokhlov:2000}, 
$\eta'\pi$~\cite{Beladidze:1993,Ivanov:2001}, $f_{1}\pi$~\cite{Kuhn:2004,Amelin:2005}, and $\omega\pi\pi$~\cite{Amelin:2005,Lu:2005}). 
In particular the resonant nature of the $\pi_1(1600)$ in the $\rho\pi$ decay channel observed in $3\pi$ final states is highly 
disputed, see e.g.~\cite{Amelin:2005,Dzierba:2006}. 

COMPASS has started to shed new light on the puzzle of spin-exotics by the
observation of a $1^{-+}$ signal in the 2004 data, as briefly
summarised in Sec.\,\ref{subsec.2004}.
%% consistent with the famous $\pi_1(1600)$~\cite{Alekseev:2009a}, cf. Sec.\,\ref{subsec.2004}. 
%% It shows clean phase motions with respect to other waves, confirming the resonance nature~\cite{Alekseev:2009a}, see Sec.\,\ref{subsec.2004}.            
The COMPASS two-stage spectrometer~\cite{compass:2007} features high resolution electromagnetic calorimetry in both stages, crucial for 
decay channels involving $\pi^{0}$, $\eta$ or $\eta'$. A Ring Imaging Cherenkov (RICH) detector allows for final state particle
identification (PID). 
%% The good separation of pions from kaons enables the study of kaonic
%% final states. 
Since 2008, two Cherenkov Differential counters with Achromatic Ring 
focus (CEDAR) upstream of the target are used to identify the incoming 
beam particle. Not only production of strangeness with the pion beam
can thus be studied but also kaon diffraction, tagging the kaon
contribution in the negative hadron beam (96\,\% $\pi^{-}$, 3.5\,\% $K^{-}$, 0.5\,\% $\bar{p}$). 
In addition to the excellent opportunity for simultaneous observation of
new states in various decay modes within the same experiment 
(Sec.\,\ref{subsec.2008}, \ref{subsec.2008b}), the data include 
subsets with different beam projectiles ($\pi^{\pm}, K^{\pm}, p$) and 
targets (H$_2$, Ni, W, and Pb), allowing for systematic studies not
only of diffractive and central production but also Primakoff 
reactions~\cite{grabmueller:2010} and even baryon spectroscopy~\cite{austregesilo:2010}.  
\vspace{-0.3cm}
\section{First results from diffractive dissociation}
\vspace{-0.3cm}
\begin{figure}[tp!]
\vspace{-0.6cm}
  \begin{minipage}[h]{.49\textwidth}
    \begin{center}
      \includegraphics[clip,trim= 3 4 22 5,width=0.9\linewidth,
       angle=0]{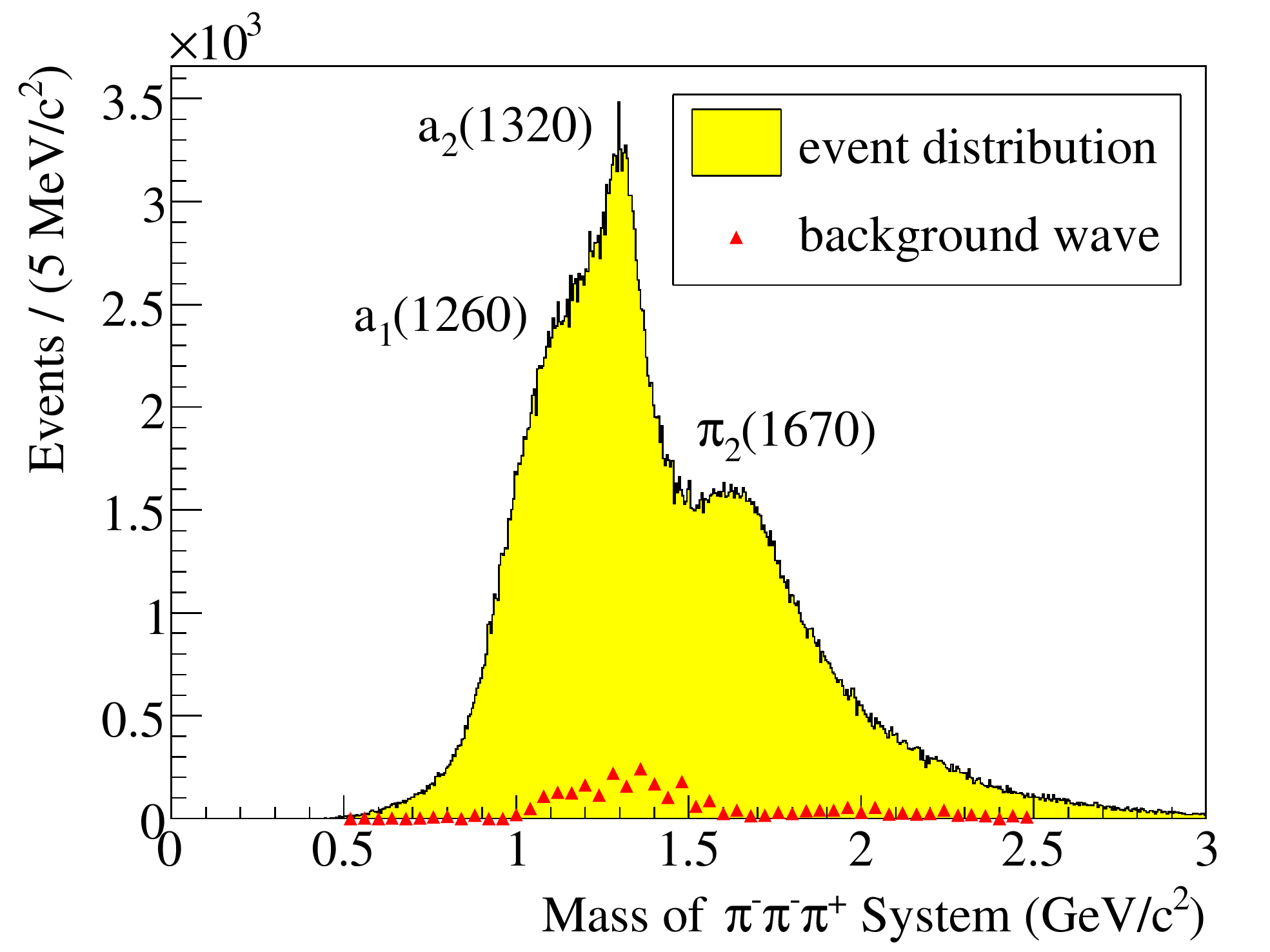}
    \end{center}
  \end{minipage}
  \hfill
  \begin{minipage}[h]{.49\textwidth}
    \begin{center}
      \includegraphics[clip,trim= 24 15 10 360,width=0.95\linewidth,
     angle=0]{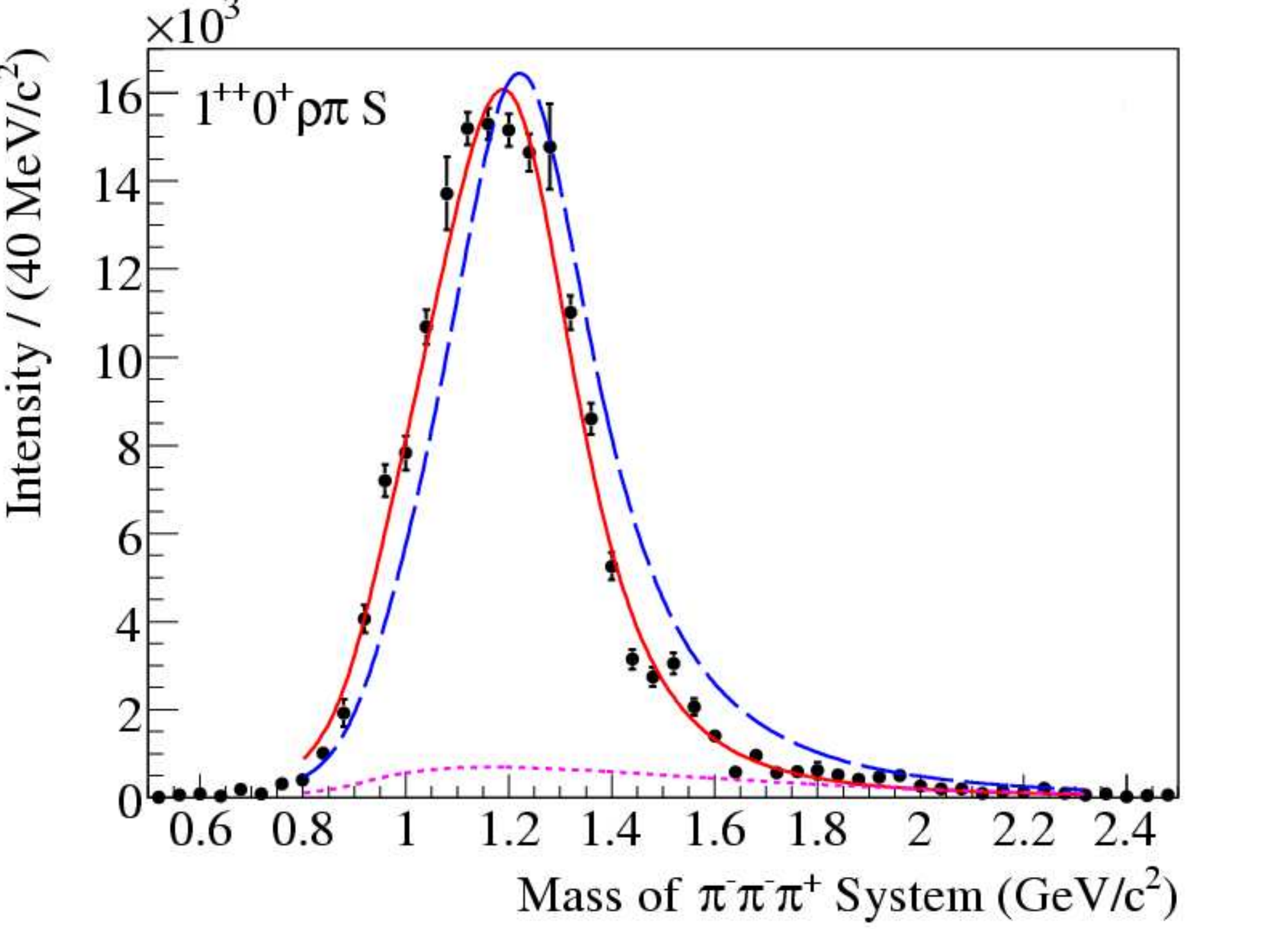}
    \end{center}
  \end{minipage}
\begin{minipage}[h]{.49\textwidth}
    \begin{center}
      \includegraphics[clip,trim= 4 15 30 360,width=0.95\linewidth,
	angle=0]{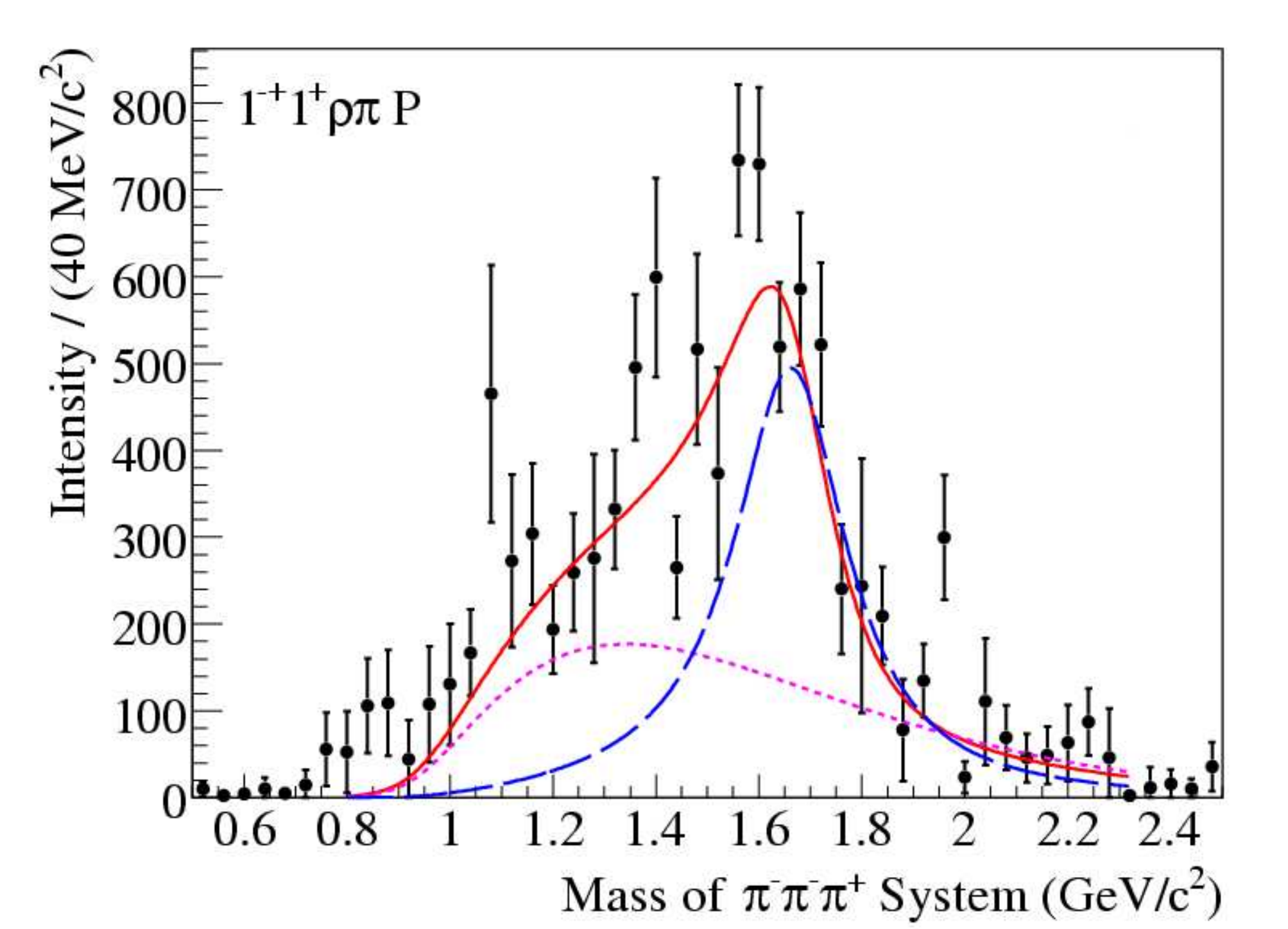}
    \end{center}
  \end{minipage}
  \hfill
  \begin{minipage}[h]{.49\textwidth}
    \begin{center}
      \includegraphics[clip,trim= 24 15 10 360,width=0.95\linewidth,
     angle=0]{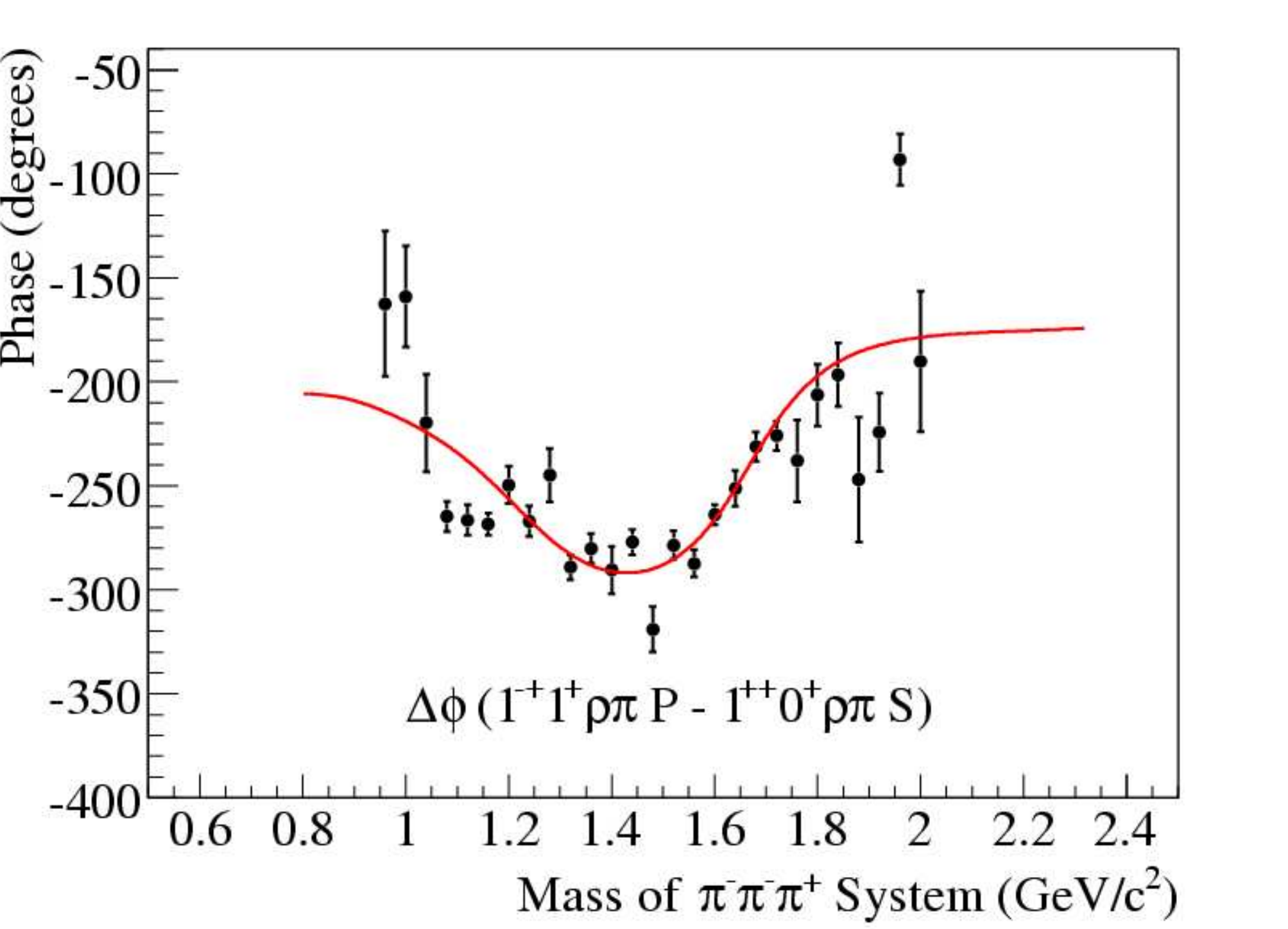}
    \end{center}
  \end{minipage}
      \caption{Top: {\it (Left)} $3\pi$ invariant mass, the most prominent resonances are indicated. {\it (Right)} Fitted intensity for the $a_{\rm 1}(1260)$: Intensity of $1^{\rm ++} 0^{\rm +}[\rho^{-} \pi] S$ wave. 
Bottom: {\it (Left)} PWA fits of the exotic $1^{\rm -+} 1^{\rm +}[\rho \pi] P$ wave. {\it (Right)} Phase motion of the exotic $1^{\rm -+}1^{\rm +}$ versus $1^{\rm ++} 0^{\rm +}$ wave.}
      \label{fig:PWA2004}
\vspace{-0.5cm}
\end{figure}
First 3$\pi$ final states are discussed in Secs.\,\ref{subsec.2004},\ref{subsec.2008}), and then kaonic final states in Sec.\,\ref{subsec.2008b}.
\subsection{Observation of a $J^{PC} = 1^{-+}$ exotic resonance -- 2004 data {\normalsize\it (negative beam, Pb target)}}
\label{subsec.2004}
\vspace{-0.2cm}
%%\paragraph{Observation of a $J^{PC} = 1^{-+}$ exotic resonance -- 2004 data {\normalsize\it (negative beam, Pb target)}}
%%~\\
%%\noindent
The quantum numbers spin $J$, parity $P$ and $C$-parity of the produced resonance $X^{-}$, together with 
the spin projection given by $M$ and $\epsilon$ (reflectivity), define a partial-wave $J^{PC}M^\epsilon[isobar]L$.
The partial-wave analysis (PWA) is based on the isobar model, see Fig.\ref{fig:diffrProd_Spectro} {\it (right)}. 
The resonance $X^{-}$ decays via an intermediate di-pion resonance (the isobar), accompanied by a so-called bachelor pion.
%, with relative orbital angular momentum $L$. 
The PWA method consists of two steps. First, a mass-independent fit is performed on the data binned into 40\,MeV/c$^2$ wide 
mass intervals, no assumption on the resonance structure of the $3\pi$ system is made at this level. 
A total set of 42 waves including an isotropic background wave is fitted to the data using an extended maximum likelihood method, 
which takes into account acceptance corrections. 
\begin{figure}[tp!]
  \begin{minipage}[h]{.49\textwidth}
    \begin{center}
\vspace{-0.7cm}
      \includegraphics[clip,trim= 3 4 22 5,width=0.9\linewidth,
       angle=0]{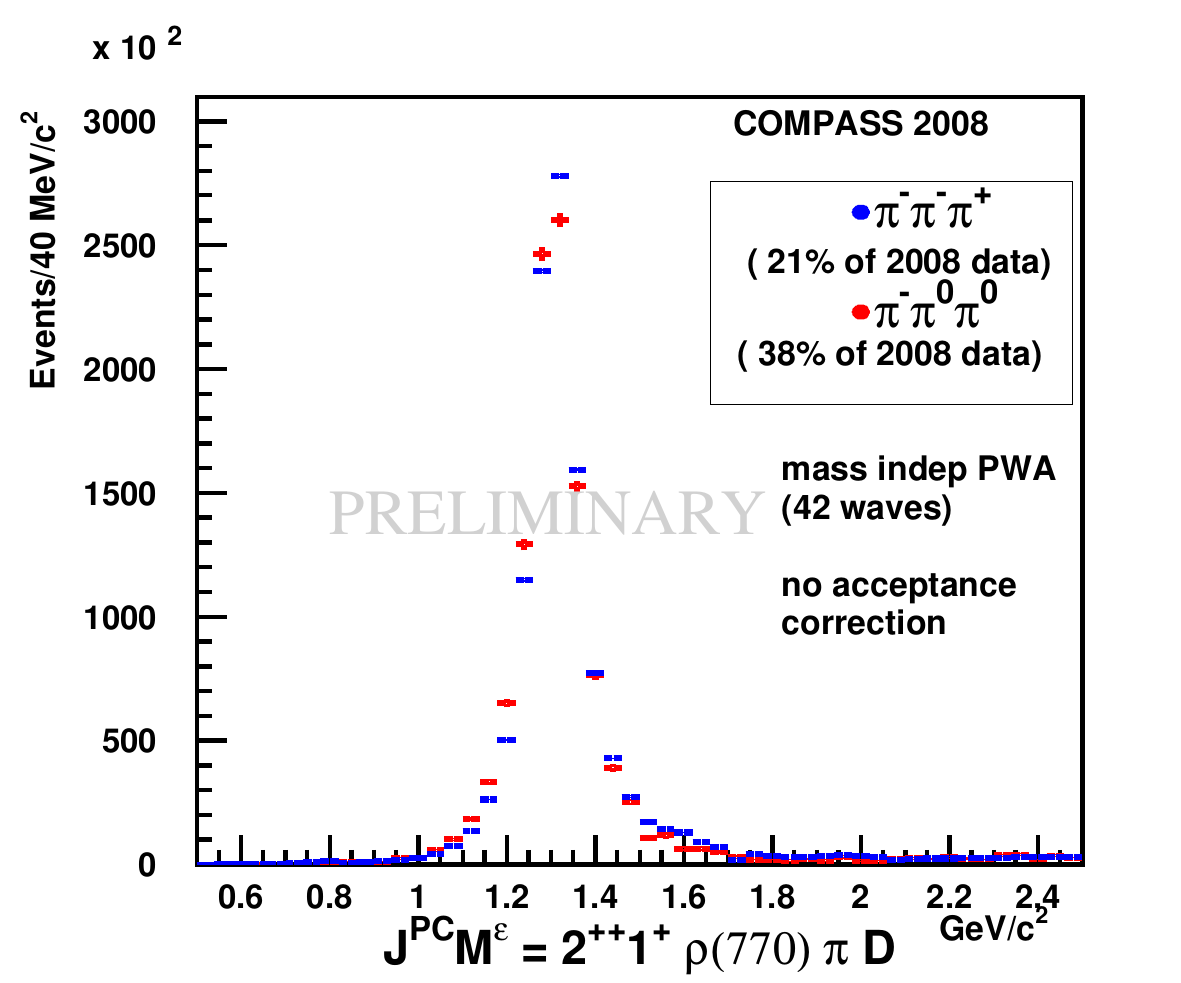}
    \end{center}
  \end{minipage}
  \hfill
  \begin{minipage}[h]{.49\textwidth}
    \begin{center}
\vspace{-0.7cm}
      \includegraphics[clip,trim= 3 4 22 5,width=0.9\linewidth,
     angle=0]{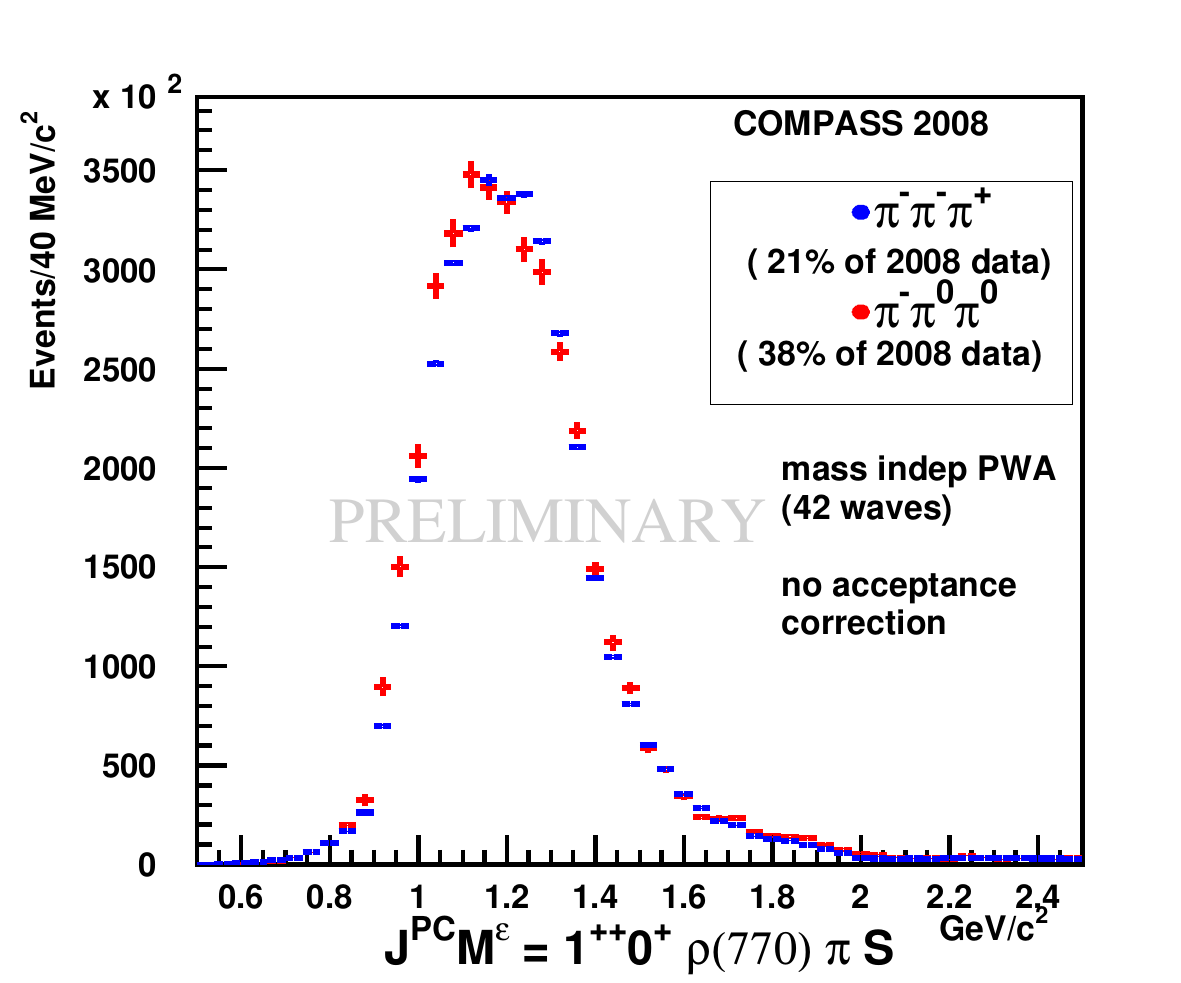}
    \end{center}
  \end{minipage}
  \begin{minipage}[h]{.49\textwidth}
    \begin{center}
\vspace{-0.4cm}
      \includegraphics[clip,trim= 3 2 22 5,width=0.9\linewidth,
	angle=0]{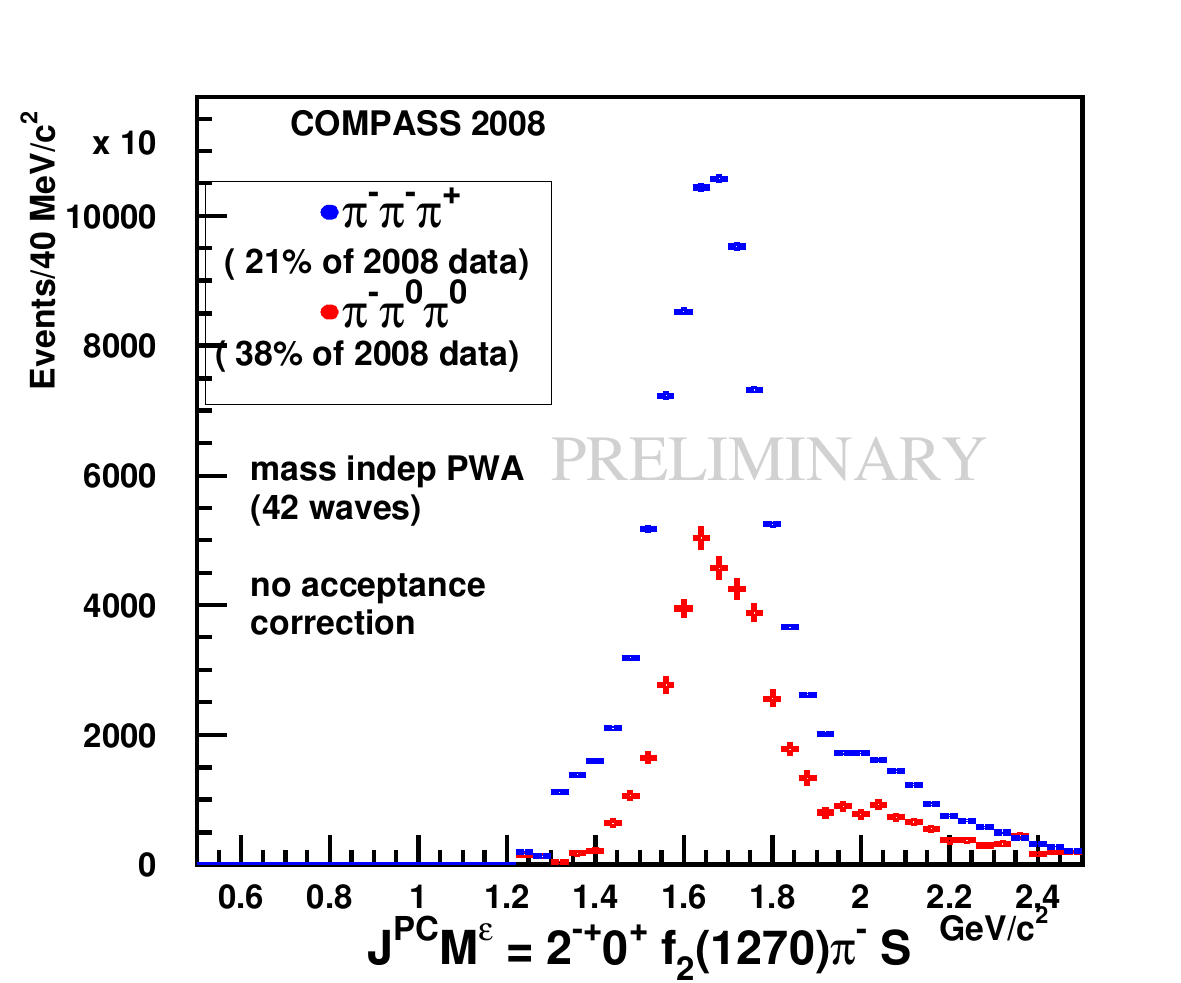}
    \end{center}
  \end{minipage}
  \hfill
  \begin{minipage}[h]{.49\textwidth}
    \begin{center}
\vspace{-0.4cm}
      \includegraphics[clip,trim= 3 2 22 5,width=0.9\linewidth,
     angle=0]{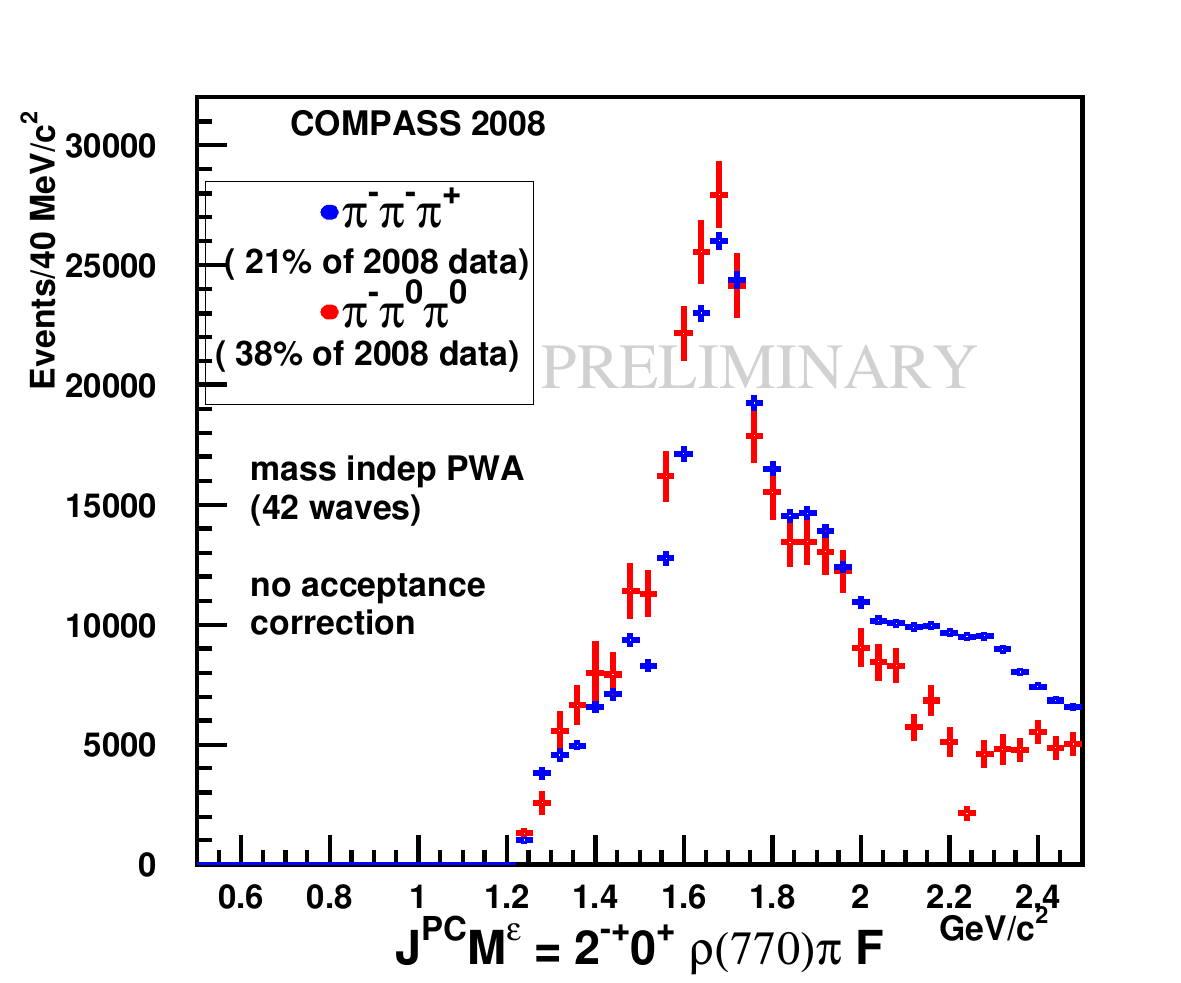}
    \end{center}
  \end{minipage}
      \caption{Comparison of PWA intensities of main waves for neutral vs. charged mode. \newline 
       {\it (Top/Left)} Intensities of the $a_{\rm 2}$ ($2^{\rm ++}1^{\rm +}$ going into $\rho^{-} \pi$ D wave) 
       used for normalisation of charged to neutral mode. Top/Right: ($a_{\rm 1}$) $1^{\rm ++} 0^{\rm +}$ 
       into $\rho^{-} \pi$ D wave. 
       {\it (Bottom/Left)} ($\pi_{\rm 2}$) $2^{\rm -+} 0^{\rm +}$ into $f_{\rm 2}$(1270) $\pi$ S wave. 
       {\it (Bottom/Right)} ($\pi_{\rm 2}$) $2^{\rm -+} 0^{\rm +}$ into $\rho^{-} \pi$ F wave.}
      \label{fig:PWA2008}
\vspace{-0.5cm}
\end{figure}
Subsequently, the mass-dependent fit is applied simultaneously to the six (main) waves out of the result from the first step using a $\chi^2$ 
minimisation. 
The mass dependence is parameterised by relativistic Breit-Wigners (BW) and coherent background, if present. 
The employed parameterisation of the spin density matrix has a rank of two, accounting for spin-flip and spin-non-flip amplitudes 
at the baryon vertex. 
Fig.\,\ref{fig:PWA2004} shows the intensity of the $1^{++}$ wave with the well-established $a_1(1260)$ and that of the spin-exotic $1^{-+}$ wave 
as well as the phase difference $\Delta\Phi$ between the two, for the fits of the other four waves, see \cite{Alekseev:2009a}. 
The black data points represent the mass-independent fit, whereas the mass-dependent one is overlayed as solid line. The separation into background 
(dotted) and BW (dashed) is plotted where applicable. Especially the resonant nature of the exotic $1^{-+}1^{+}[\rho\pi] P$ wave was questioned in previous observations, whereas our data show a clear and rapid phase motion.
%, see Fig.\ref{fig:PWA2004}.
%, and confirms the resonant nature. 
Our result of a mass of $1660\pm 10^{+0}_{-64}$\,MeV/c$^2$ and a width of $269\pm 21^{+42}_{-64}$\,MeV/c$^2$ is consistent with the 
$\pi_1(1600)$~\cite{Alekseev:2009a} already reported in the past but still controversially discussed.

\subsection{First comparison of neutral versus charged mode  -- 2008 data {\normalsize\it (negative beam, H$_2$ target)}}
\label{subsec.2008}
\vspace{-0.2cm}
%%\paragraph{First comparison of neutral versus charged mode  -- 2008 data {\normalsize\it (negative beam, H$_2$ target)}}
%%~\\
%%\noindent
An important cross-check of all analyses is the test for isospin symmetry in the observed spectra.
The $\rho\pi$ decay channel of the $\pi_1(1600)$ for example, can be studied in two modes of 3$\pi$ 
final states, $\pi^{-}\pi^{+}\pi^{-}$ (charged) and $\pi^{-}\pi^{0}\pi^{0}$ (neutral), respectively.
The relative contribution should follow isospin conservation, depending on the underlying isobars, as
it is shown in Fig.\,\ref{fig:PWA2008}. 
A first partial-wave analysis of main waves in diffractively produced 3$\pi$ events has been performed for both modes, 
applying the same model as for the 2004 result. The analysis shown here is based on a data sample
increased in statistics by nearly a factor of five as compared to~\cite{nerling:2009}. The 2008 results have not yet 
been acceptance corrected, however, the effect is estimated to be up to $\sim 20\,\%$, for details see~\cite{nerling:2009}. 
The wave intensities shown are normalised to the well-known $\rho\pi$ decay 
of the $a_2(1320)$ to compensate for the different statistics presently analysed for the different modes, thus making them comparable. We find similar intensities for the $\rho\pi$ decays in both modes, whereas a suppression factor of two is observed for the waves decaying into $f_2\pi$ in the neutral mode -- as expected due to 
the Clebsch-Gordon coefficients. 

Further ongoing analyses of neutral channels cover $\pi^{-}\eta$ and $\pi^{-}\eta\eta$ final states (search for the 
$\pi_1(1400)$ and lightest $0^{++}$ glueball candidate) as well as $\pi^{-}\pi^{-}\pi^{+}\pi^{0}$,
$\pi^{-}\pi^{-}\pi^{+}\eta$ and $\pi^{-}\pi^{-}\pi^{+}\pi^{0}\pi^{0}$ final states (accessible isobars: 
$f_1, b_1, \eta, \eta', \omega$). For all these channels, COMPASS has recorded significantly higher statistics 
with respect to previous experiments, covering all spin-exotic meson decay channels in the light quark sector 
reported in the past.     

\subsection{First glimpse on kaonic final states  -- 2008 data {\normalsize\it (negative beam, H$_2$ target)}}
\label{subsec.2008b}
\vspace{-0.2cm}
%%\noindent
%%\paragraph{First glimpse on kaonic final states  -- 2008 data {\normalsize\it (negative beam, H$_2$ target)}}
%%~\\
\begin{figure}[tp!]
  \begin{minipage}[h]{.49\textwidth}
    \begin{center}
\vspace{-0.5cm}
      \includegraphics[clip,trim= 10 5 25 15,width=1.0\linewidth,
	angle=0]{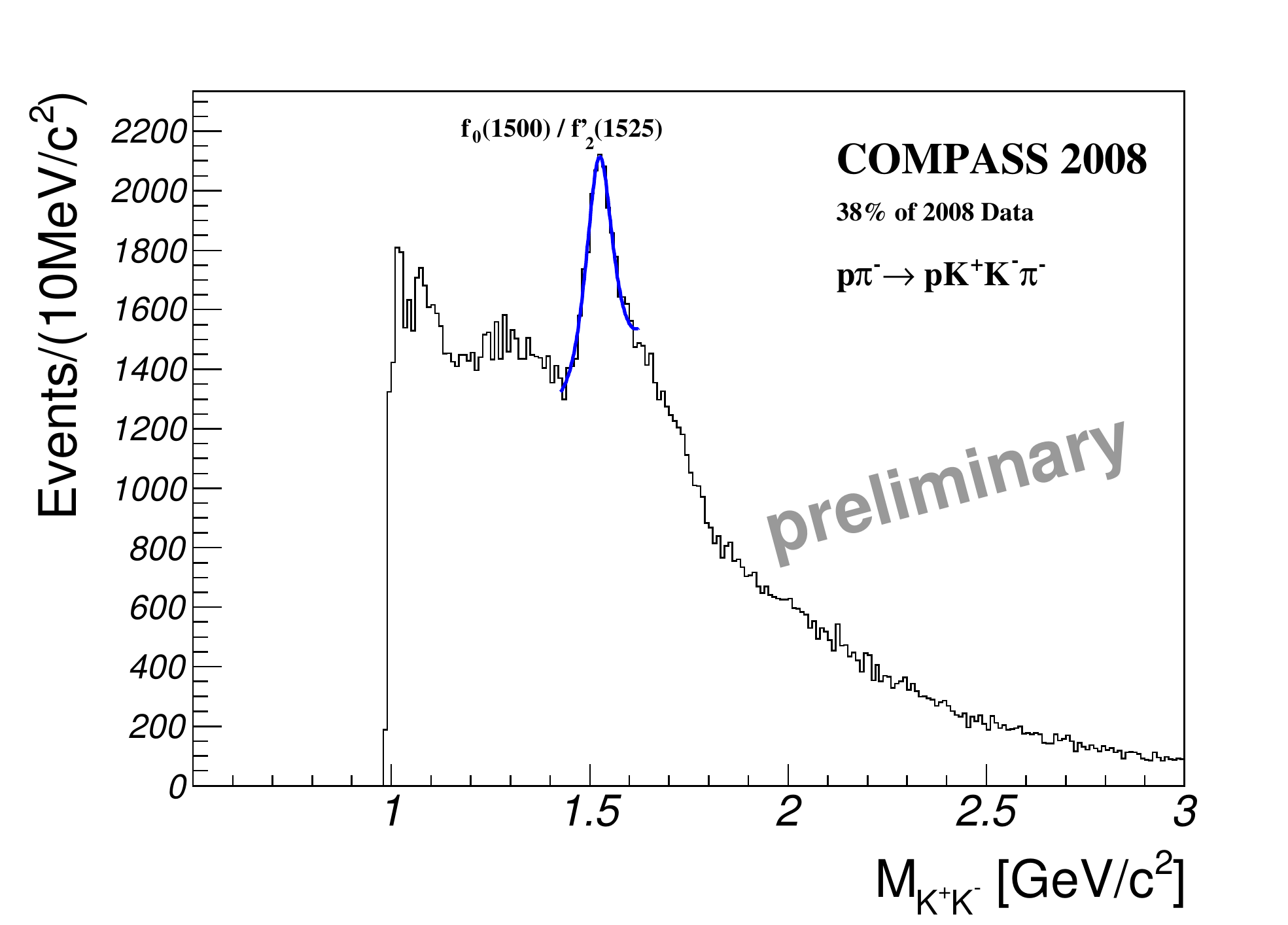}
    \end{center}
  \end{minipage}
  \hfill
  \begin{minipage}[h]{.49\textwidth}
    \begin{center}
\vspace{-0.5cm}
       \includegraphics[clip,trim= 10 0 25 20,width=1.0\linewidth,
     angle=0]{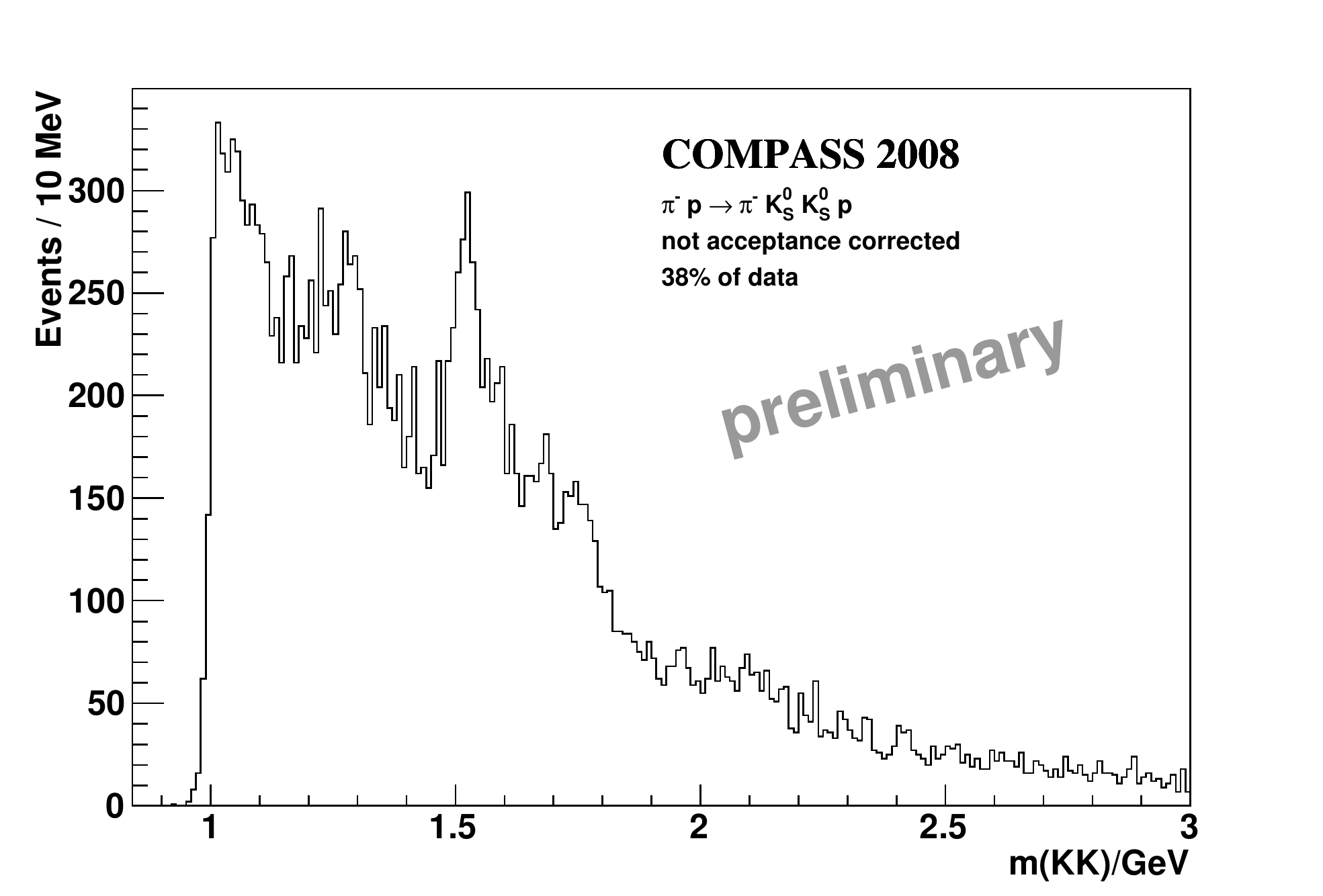}
    \end{center}
  \end{minipage}
      \caption{Invariant mass spectra of the diffractively produced $(K\bar{K}\pi)^{-}$ systems with the $\pi^{-}$ beam on the liquid hydrogen target: 
	{\it (Left)} $K^+K^-$ (with $p_{K^-} \le 30$\,GeV/c) {\it (Right)} $K^0_sK^0_s$.}
      \label{fig:Kaons}
\vspace{-0.5cm}
\end{figure}
Final states including strange particles are interesting for both, glueball search in central production 
as well as for diffractively produced hybrids. Fig.\,\ref{fig:Kaons} shows the $K\bar{K}$ subsystems 
out of the $(K\bar{K}\pi)^{-}$ system, again for two different modes: $K^{+}K^{-}\pi^{-}$ (charged mode) and 
$K^{0}_s K^{0}_s\pi^{-}$ (neutral mode) final states, respectively. 
In both cases, the spectra show a clear structure around the expected $f_0(1500)$. The CEDARs were used to 
anti-tag the kaons in the beam, and the final state kaons are identified using the RICH detector and the well resolved 
$V^{0}$ secondary vertex, respectively; for details on the event selections, see~\cite{tobi:2009}. 
\begin{figure}[bp!]
  \begin{minipage}[h]{.49\textwidth}
    \begin{center}
\vspace{-0.5cm}
      \includegraphics[clip,trim= 10 5 25 15,width=1.0\linewidth,
	angle=0]{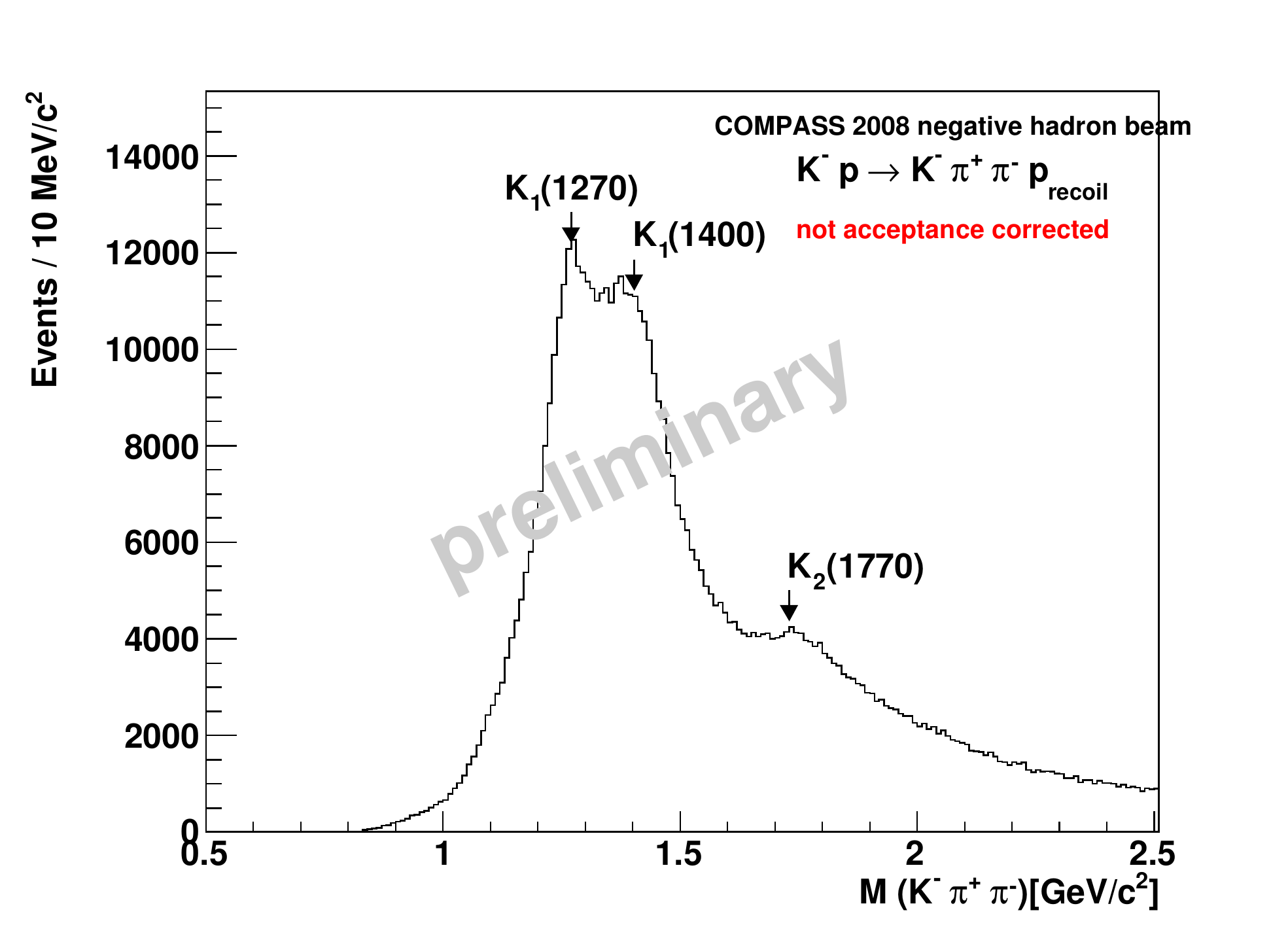}
    \end{center}
  \end{minipage}
  \hfill
  \begin{minipage}[h]{.49\textwidth}
    \begin{center}
\vspace{-0.5cm}
      \includegraphics[clip,trim= 10 0 25 20,width=1.0\linewidth,
     angle=0]{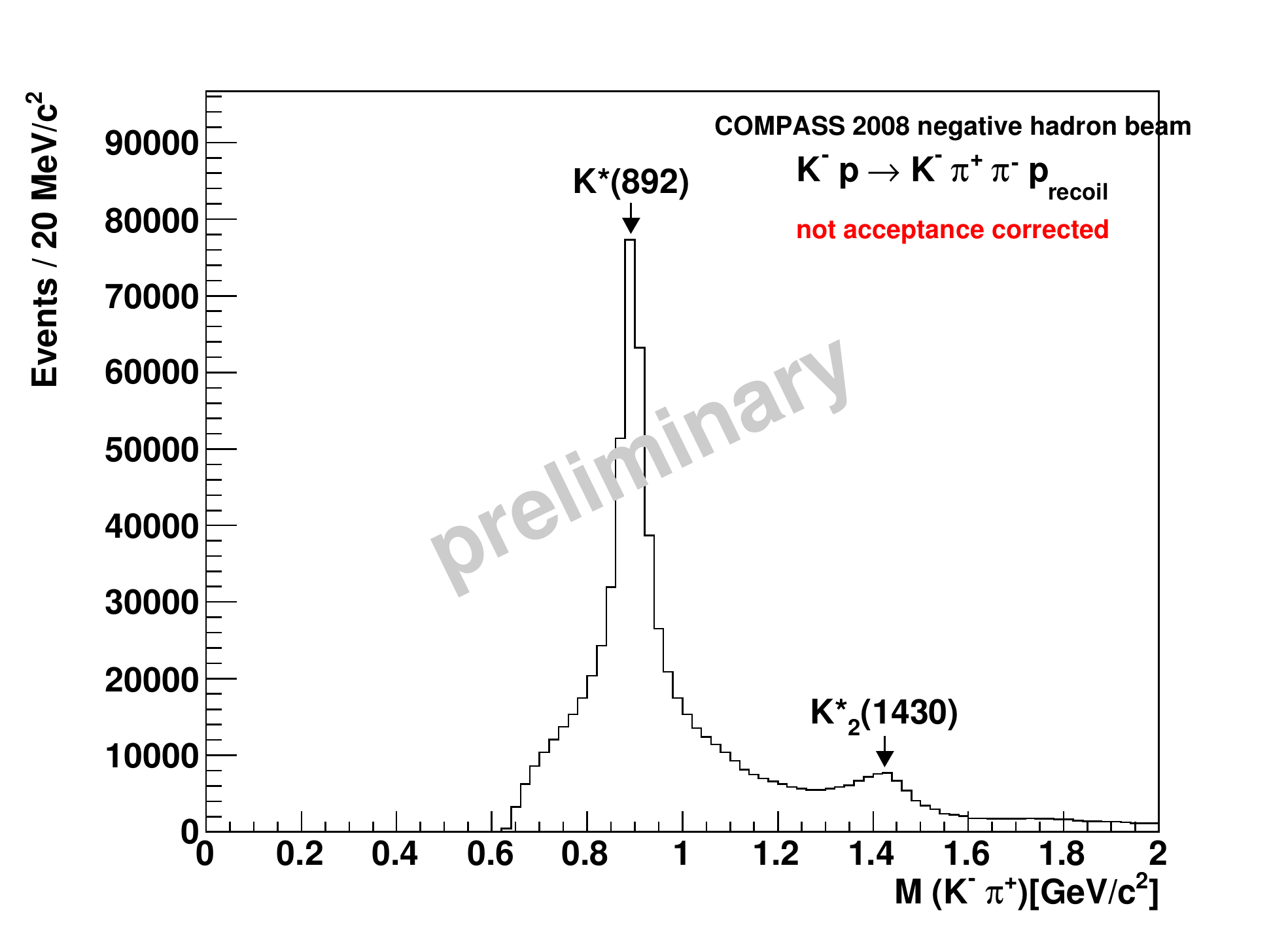}
    \end{center}
  \end{minipage}
      \caption{Kaon diffraction into $K^{-}\pi^{+}\pi^{-}$ final states:
	{\it (Left)} Invariant mass of the total diffractively produced system. 
	{\it (Right)} Invariant mass of the subsystem $K^{-}\pi^{+}$.}
      \label{fig:Kaons_b}
\vspace{-0.5cm}
\end{figure}

Further ongoing analyses of kaonic final states cover the $(K\bar{K}\pi\pi)^{-}$ system diffractively produced with the $\pi^{-}$ 
beam as well as kaon diffraction into $K^{-}\pi^{+}\pi^{-}$ final states using the beam kaons. In both cases, the RICH detector is used for 
final state PID and the CEDARs for identifying the beam particle. 
Fig.\,\ref{fig:Kaons_b} shows mass spectra for the study of kaon diffraction into $K^{-}\pi^{+}\pi^{-}$ final 
states. The total mass spectrum and the $K^{-}\pi^{+}$ show the prominent, well-known resonances as expected, similar
as observed by WA03. The statistics collected in 2008/09 exceeds the one from WA03 by about a factor of five, for details, 
see \cite{promme:2009} and references therein.
The PWA of all kaonic final states mentioned in this paper are under preparation.

\vspace{-0.3cm}
\section{Summary \& conclusions}
\vspace{-0.3cm}
COMPASS has taken data with high-intensity, negatively as well as positively charged hadron beams 
($\pi^{\pm},K^{\pm},p$) on nuclear and liquid hydrogen targets. 
The data sample newly taken in 2008/09 exceeds the world data by a factor of 5-100, depending on 
the given final state. The COMPASS data sets allow to address open issues in light-mesons spectroscopy 
with good accuracy, extending the kinematics region to higher masses beyond 2\,GeV/c$^2$.   

%%\vspace{-0.3cm}
\section*{Acknowledgements}
\vspace{-0.3cm}
This work is supported by the BMBF (Germany), particularly the ``Nutzungsinitiative CERN''.

%%\vspace{-0.3cm}

\end{document}